\documentclass[acmsmall,screen]{acmart}
\setcopyright{none}
\copyrightyear{2025}
\acmYear{2025}
\acmDOI{}

\usepackage{fancyhdr}
\usepackage{subcaption}

\usepackage{tikz}
\usepackage{pgfplots}

\AtBeginDocument{%
  }

\begin{document}

\title{concurrentKanren: miniKanren for parallel execution}

\author{Sjoerd Dost}
\email{deosjr@gmail.com}

\settopmatter{printacmref=false}
\settopmatter{printfolios=true}
\renewcommand\footnotetextcopyrightpermission[1]{}
\pagestyle{fancy}
\fancyfoot{}
\fancyfoot[R]{miniKanren'25}
\fancypagestyle{firstfancy}{
  \fancyhead{}
  \fancyhead[R]{miniKanren'25}
  \fancyfoot{}
}
\makeatletter
\let\@authorsaddresses\@empty
\makeatother

\begin{abstract}
Concurrent logic programming predates miniKanren, but concurrent implementations of miniKanren have remained largely unexplored.
 In this work we present a parallel implementation of miniKanren in Go, demonstrating its feasibility and potential for performance improvements.
Our approach leverages implicit parallelism allowing legacy programs to benefit from parallel execution.
We discuss implementation strategies and evaluate the impact of parallelism, laying groundwork for future language-agnostic models.
\end{abstract}

\maketitle
\thispagestyle{firstfancy}

\section{Introduction}

Parallelization of logic programming languages has been actively studied for a long while, resulting in the family of concurrent logic programming languages \cite{Shapiro1989}.
miniKanren is a logic programming language with functional programming principles influencing its design.
This makes it well suited for parallelization, as noted in \cite{WillOnProlog}.
To our knowledge no concurrent implementation exists, though \cite{Assylkhanova2023} is a step in that direction, running multiple queries in parallel.
A concurrent implementation could speed up execution when ran in parallel, if overhead can be managed.
Consider for example the following query:
\begin{verbatim}
    (run* (x) (disj (equalo x 5) (equalo x 6)))
\end{verbatim}
The expected outcome is a stream of two results, one per subgoal: \texttt{x} can either be equal to \texttt{5} or to \texttt{6}.
Neither outcome of the subgoals of the disjunction depends on the other.
Testing whether \texttt{x} is equal to \texttt{5} or \texttt{6} is done in the same evaluation context, which is immutable by default.
It is therefore tempting to think these two evaluations can happen concurrently.\\

Concurrent logic programming languages have shown that this intuition and similar ideas can lead to interesting and useful new results.
We believe that while miniKanren might be `trivially parallelizable' in theory \cite{WillOnProlog}, getting to a practical implementation is not trivial.
In this paper we introduce \emph{concurrentKanren}, a concurrent implementation of miniKanren.
Our aim is to modify the language as little as needed, such that it would be feasible to embed a miniKanren with concurrency support in any language.
This paper makes the following contributions:
\begin{itemize}
\item We implement miniKanren with support for parallel evaluation using implicit parallelism. We discuss the conceptual changes to $\mu$Kanren in Section \ref{idea} as they arise from a naive implementation in Go. The final Go implementation and its specifics are discussed in Section \ref{golang}.
\item We introduce \emph{disj\_conc}, a concurrent generalized disjunction goal with improved fairness of evaluation (Section \ref{disjconc}). 
\item We introduce \emph{conj\_sce}, a concurrent conjunction goal that uses short-circuit evaluation to handle diverging subgoals, reducing impact of goal order (Section \ref{conjsce}).
\item We benchmark two implementations against a single-threaded baseline: a message-passing and a worker pool-based version (Section \ref{results}). We also compare against the current single-threaded standard, \emph{faster-minikanren} \cite{fasterminikanren}.
\end{itemize}

\section{miniKanren} \label{minikanren}

This section outlines the semantics of miniKanren relevant to our implementation.
miniKanren refers to a family of lightweight logic languages, frequently embedded as a domain-specific language.
Initially introduced in \cite{ReasonedSchemer} and defined extensively in \cite{WillThesis}, the core of miniKanren is small enough to be easily embedded in many different languages.
We will take the minimal language from \cite{Hemann2013} embedded in Scheme as our starting point and canonical implementation.
We have already seen an example query in the introduction. When we execute this Scheme code, we get:

\begin{verbatim}
    > (run* (x) (disj (equalo x 5) (equalo x 6)))
    (5 6)
\end{verbatim}

If x can be 5 or 6, then there are two valid states: one in which \texttt{x} equals \texttt{5}, and the other in which \texttt{x} equals \texttt{6}.
The miniKanren model deals with \emph{expressions}, which can be any expression of the host language that is supported, and \emph{variables} (which are also valid expressions) that can be associated with an expression.
A variable associated with an expression is said to be bound to that expression.
The \emph{value} of a variable is the non-variable expression that it is transitively bound to in the state, if any.

The result of a query is a list of states, where each \emph{state} is a substitution that is compatible with the query, mapping variables to expressions.
A query is composed of \texttt{run*} or \texttt{run} (which takes an extra argument limiting the number of results), a list of free (unbound) variables, and a \emph{goal} in which those variables are introduced.
Goals are functions that take a state and return a \emph{stream} of states. In $\mu$Kanren, streams are implemented as cons lists. 
We can take any state as our starting point, but by default we use the empty state (i.e. no variables have been bound) in queries that we \texttt{run}.
Core miniKanren is built on four primitive goals: \emph{equalo} testing equality, \emph{disj} for disjunction, \emph{conj} for conjunction, and \emph{call/fresh} for introducing new variables.\\

We can reason about infinite streams and ask for a bounded number of results:

\begin{verbatim}
    > (define (fives x) (disj (equalo x 5) (delay (fives x))))
    > (define (sixes x) (disj (equalo x 6) (delay (sixes x))))
    > (run 9 (x) (disj (fives x) (sixes x)))
    (5 6 5 6 5 6 5 6 5)
\end{verbatim}

The definition of \texttt{fives} and \texttt{sixes} is recursive, with the \texttt{delay} construct serving as a temporary stopping point preventing stack overflow from infinite evaluation.
This is made possible by allowing streams to either be \emph{mature} (a fully evaluated list of states) or \emph{immature} (a \emph{thunk}, a procedure with no arguments).
When disjunction encounters an immature stream from a subgoal it attempts to mature it to get a result, but also prioritises results from the other subgoal's stream.
This \emph{binary trampolining} allows for interleaving of infinite streams.

\section{Parallelization}

The standard miniKanren evaluation model is single-threaded.
Our goal is to make use of modern multi-core processors to speed up search.
The first design decision we have to make is whether to bring parallel execution under the programmer's control.
This is the distinction between \emph{explicit} and \emph{implicit} parallelism.
Explicit parallelism introduces syntax to denote a goal as executing its subgoals concurrently if possible.
This gives fine-grained control to the programmer, but also means that they have to be aware of the new syntax and when to use it.
Implicit parallelism attempts to hide this overhead from the user.
Our focus on implicit parallelism avoids modifying the language syntax and allows existing miniKanren code to potentially benefit from improved runtime performance.

The two main opportunities for parallelism in logic programming are disjunction (\emph{OR-parallelism}) and conjunction (\emph{AND-parallelism}).
Disjunction has the benefit that the outcome of each of its subgoals does not at all depend on the outcome of the other: they can be evaluated independently.
This stands in stark contrast to conjunction, in which the evaluation of the second goal takes results from the first as its input.
We consider OR-parallelism therefore a stronger candidate for potential speed gains.
Both forms of parallelism can bring us benefit however, as we describe in this next section.

\subsection{Disjunction}

In a disjunction results from both subgoals are completely independent, and merging results can be done without inspecting them.
This means that instead of evaluating and interleaving results one at a time, we could do both at the same time.
This raises a problem: results are no longer guaranteed to come in in order.
Changing the order of results would potentially break existing miniKanren programs, which we do not want to do.
If we manage to solve that though, the approach could easily extend to disjunctions over more than two goals.

Those disjunctions are usually written as a macro (\texttt{disj+}), nesting disjunction goals \cite{Hemann2013}.
This implements binary trampolining, but introduces a problem with logical ordering of results.
Consider this example, building upon the ones in Section \ref{minikanren}:

\begin{verbatim}
    > (define (sevens x) (disj (equalo x 7) (delay (sevens x))))
    > (run 9 (x) (disj+ (fives x) (sixes x) (sevens x)))
    (5 6 5 7 5 6 5 7 5)
\end{verbatim}

In this example we would expect three digits alternating, but instead \texttt{5} is repeated more often.
The \texttt{disj+} macro expands into nested binary disjunctions, creating a right-branching binary tree.
This introduces a structural bias: leftmost goals are evaluated earlier and more frequently.
If all streams feeding into a disjunction are concurrently producing results anyways, solving this problem reduces to interleaving the results fairly.

\subsection{Conjunction}

In conjunction, the second subgoal depends on bindings produced by the first. Parallel execution would require inter-goal communication or speculative evaluation.
Intuitively we can understand this as both child goals introducing constraints that the resulting output stream needs to respect.
In the case of a disjunction we are looking for the set of results that satisfy either constraint set; for conjunction we are looking for results that satisfy both.
This means that we do not expect a lot of speed gains from optimising conjunction directly.
However, there is a problem with conjunction over infinite streams that could potentially benefit:

\begin{verbatim}
    > (define (failo x) (equalo 1 2))
    > (run* (x) (failo x))     ; evaluates to ()
\end{verbatim}

The \texttt{failo} goal will never succeed. A conjunction that includes it as a subgoal will therefore also never succeed.
Since we evaluate the two subgoals of a conjunction in order, this gives us unexpected behaviour when combined with an infinite stream: 

\begin{verbatim}
    > (run* (x) (conj (failo x) (fives x)))         ; evaluates to ()
    > (run* (x) (conj (fives x) (failo x)))         ; diverges
\end{verbatim}

From a logical standpoint the subgoal order should not matter but here it makes all the difference.
As a solution we could run both subgoals in parallel, and halt the entire evaluation of conjunction if one is found to be unproductive.
This is equivalent to short-circuit evaluation as seen in many programming languages.
Although running both subgoals in parallel may be inefficient, it restores logical symmetry: the result no longer depends on goal order.

\section{Parallel Evaluation Model} \label{idea} 

This section describes the concurrentKanren model through code examples, starting from a very naive approach and working through challenges as they arise.
Our model is inspired by the actor paradigm, where each goal is treated as an independent process managing its own stream of results.
We focus on how streams, disjunctions, and delays are reinterpreted in this concurrent setting.
At the end we introduce two new goals, one for disjunction and one for conjunction.
We choose the Go language for our code examples as it is the language of our reference implementation.
Implementation-specific details and discussions are deferred to Section~\ref{golang}.

\subsection{Search nodes as actors} \label{searchnodes}

In the miniKanren evaluation model, each goal can be seen as a node in a search graph evaluating the state space of potential solutions. 
Each query has a corresponding graph, a tree structure in which nodes have a single parent and a number of children.
The first change we make to the standard model is to represent each goal as its own process with its own stream to manage.
Processes can create other processes as part of their execution.

Streams can be infinite so we will manage them lazily, evaluating up until the next result in response to a request for more answers.
This behaviour is similar to a single iteration of miniKanren's \texttt{take} function, but happening locally in the search tree.\\

The result of executing a goal constructor such as \texttt{disj} is a function encapsulating the inputs in a closure.
Each invocation of this goal function creates an output stream and a process to maintain it.
We note that each process will only ever get requests from a single source (its parent), and vice-versa will only ever request results from the child processes it has created.
For now, whenever we request a result from a stream we will immediately follow that up with a call to read the response from that stream.
This closely mimics the normal miniKanren evaluation model in which this is equivalent to function calls.
Later we will relax this constraint.

The \texttt{disj} goal constructor demonstrates a typical goal in concurrentKanren:\\
\begin{verbatim}
    func disj(g1, g2 goal) goal {									// construct a goal
        return func(st state) stream { 					// goal is function of state -> stream
            str := newStream()												// create a new stream		
            go mplus(str, g1(st), g2(st)) // run mplus in a new goroutine
            return str																				// return stream handle
        }
    }
\end{verbatim}

The thread that is evaluating \texttt{disj} creates a goal function capturing goals $g_1$ and $g_2$.
When passed a state, it creates a new stream handle and returns it.
Subgoal evaluation also happens in the calling thread, creating two new streams for the subgoals of the disjunction.
The \texttt{mplus} function is run in a new goroutine and will handle all requests for answers.
A first approximation of $\mu$Kanren's \texttt{mplus} would be:\\
\begin{verbatim}
    func mplus(str, str1, str2 stream) {
        str.getRequest()														// wait for a request
        str1.requestResult()										// request a result from stream1
        st, ok := str1.getResult()				// get a result from stream1
        if !ok {
            str.forward(str2)									// forward results
            return
        }
        str.publish(res) 													// publish new result
        mplus(str, str2, str1)								// continue with interleaving
    }
\end{verbatim}

When \texttt{mplus} gets a request for more results, it asks its first subgoal for more.
If that subgoal is productive we have found an answer: it is returned to the caller and we switch subgoals, effectively interleaving results.
If no solution is found we know that the remainder of \texttt{mplus} is equivalent to the result stream of the second subgoal.
In this case we forward all results from the second subgoal to the caller from now on.

\subsection{Message definitions}

\begin{table}[ht]
\caption{Messages definitions in concurrentKanren}
\label{tab:msg}
\begin{tabular}{lll}
\toprule
Direction&Message name&Meaning\\
\midrule
Request&Request&Request a result\\
Request&Done&Signal no more request will be coming\\
Result&State&Result found, more might be available\\
Result&StateAndClose&Result found, no more available\\
Result&Close&No result, no more available\\
Result&Forward&Result may be found on other stream\\
Result&ForwardWithState&Result found, further results on other stream\\
Result&Delay&Immature stream\\
\bottomrule
\end{tabular}
\end{table}

Unlike in traditional miniKanren, where evaluation proceeds via function return values and thunks, our processes interact asynchronously.
A well-defined set of message types allows us to handle termination, laziness, and forwarding without race conditions or ambiguous control flow.
Messages fall into two broad categories: requests and results.
Request messages prompt a process to evaluate further or indicate the query is finished, while result messages represent output states, delays, forwarding information, or signals that no more results are forthcoming.
Table \ref{tab:msg} shows the full set of messages used in concurrentKanren.\\

One of the most striking features is the apparent duplication and combination of different signals.
There are multiple message types that communicate results are found, forwarding has happened, or that a stream has become unproductive.
This is motivated initially by the specifics of using Go channels (see Section {\ref{channels}), but has parallels in work on miniKanren semantics.
For example, \cite{Dmitry2019} contains operational semantics that includes different rules for goal evaluation stopping with or without a result.
Similarly, faster-minikanren makes a special case distinction for streams that have exactly one state remaining, detecting both a new result and the fact that the stream will afterwards no longer be productive.

\subsection{Forwarding results}

Since we are sending messages between processes running in different threads, we can no longer rely on the control flow of the standard miniKanren model.
Specifically, when \texttt{mplus} finds its first child unproductive and returns the second, it returns a full stream that the caller can continue to evaluate.
In our model so far we have said that instead we will simply forward messages from one to the other.
This of course is inefficient.
Instead of leaving a forwarding node in our search graph, we could connect parent with grandchild and replace the node altogether.
Whenever a forward message is received, the receiver swaps the stream reference it has requested a result from with the new reference the message contains.
This does mean that we have to keep track of who sends a request, because a stream's parent node can now change.
Our \texttt{mplus} now looks like this:\\

\begin{verbatim}
    func mplus(str, str1, str2 stream) {
        parent := str.getRequest()					// wait for a request
        str.requestResult(str1)								// request a result from stream1
        st, ok := str.getResult()						// get a result (from stream1)
        if !ok {
            str.forward(parent, str2)		// forward results
            return
        }
        str.publish(parent, res) 						// publish new result
        mplus(str, str2, str1)									// continue with interleaving
    }
\end{verbatim}

\subsection{Delaying evaluation}

In miniKanren, delay is used to deal with potentially infinite streams.
When first evaluated, the stream returns a thunk so that the evaluation flow continues only when we evaluate again.
All thunks are passed upwards until they reach the top level, where a \texttt{pull} procedure resolves immature streams.
This not only seems inefficient from a message-passing point of view, but it is also problematic when \texttt{pull} is the only process that can continue evaluation.
If we imagine many delayed streams sending thunks upwards to a single process, that undoes most of the work we did in distributing computation and centralizes it again.
We instead resolve delay locally and do not pass thunks around:

\begin{verbatim}
    func delay(f func() goal) goal {
        return func(st state) stream {
            str := newStream()
            go func() {
                parent := str.getRequest()
                str.sendDelay(parent)
                parent = str.getRequest()
                str.forward(parent, f()(st))
            }()
            return str
        }
    }
\end{verbatim}

Upon first request, we send a delay signal onto the stream.
This message contains no other information than that the result is delayed.
Only when requested again do we evaluate the delayed function given the input state, and notify the parent that all further results can be found on the resulting stream.

\subsection{Generalized concurrent disjunction} \label{disjconc}

We can generalize disjunction such that a single goal becomes aware of more than two subgoals, allowing it to distribute work more efficiently.
Instead of a \texttt{disj+} macro that rewrites to nested \texttt{disj} goals with 2 subgoals each, we can write one concurrent disjunction over many.
We introduce \texttt{disj\_conc}, which maintains a local buffer of answers to give out whenever requested.
The behaviour of \texttt{disj\_conc} is as follows. When created it instantiates all of its subgoals as child processes.
Then whenever requested to give a result, it first checks whether it has any in buffer.
If the buffer is empty, it requests a new result from each of its children, then awaits their response.
These new results form the new buffer. If any child is found to be unproductive, it is removed from the child list, and if no children remain active, we return a close signal.
We could relax our earlier constraint here to expect child responses out of order, as long as we sort them before returning the buffer.
This will guarantee order to be consistent with the single-threaded version.
The main sketch of \texttt{disj\_conc} is included in the Appendix.

\subsection{Conjunction with short-circuit evaluation} \label{conjsce}

Short-circuit evaluation can halt conjunction goals with a failure in the second goal, that otherwise would wait forever on a diverging first goal.
We introduce \texttt{conj\_sce}, a form of conjunction that solves this failure mode.
When \texttt{conj\_sce} is called with a state it creates both a normal conjunction stream and a dedicated stream applying only the second goal to that state.
These streams are evaluated independently.
If we get a proper (i.e. not forwarding or delay) response from the first stream, we can use that result and stop trying to short-circuit.
If it was productive we can simply forward to it for more results if needed.
If the second stream is productive, we can similarly stop and forward.
If however the first result we get is that the second stream is unproductive, we can give up on the first stream altogether and send a close-message back to the requesting stream.
Conjunction is guaranteed to fail in this state.\\

Note that for this to work, we now \emph{have} to relax our earlier constraint on parent-child interactions, because we now rely on unknown ordering.
We cannot simply always request from the second stream first, because then we have the original problem again but in reverse!
This means that we would have to keep an inbox of messages for each stream, bringing us closer to the actor model.
The main sketch of \texttt{conj\_sce} is included in the Appendix.

\section{ConcurrentKanren in Go} \label{golang}

We implement the concurrent evaluation model described in Section \ref{idea} using the Go programming language.
Go is chosen because of its well-supported concurrency primitives and tooling around debugging programs running in parallel.
In this section, we present the key implementation strategies used to realize concurrent miniKanren in Go.
We first discuss how we used Go's concurrency primitives in our model, and then several optimisations to our initial implementation that we added later.
Our codebase can be accessed at GitHub\footnote{\url{https://github.com/deosjr/concurrentKanren}}.

\subsection{Goroutines} \label{goroutines}

Processes run on goroutines, which are lightweight threads.
Go's scheduler is built to support thousands of goroutines without too much issue.
In our model each query has a search tree in which each node is a separate process and therefore a separate goroutine.
For nontrivial queries the search tree can get quite large, and the number of goroutines similarly grows.
This works well to illustrate the extreme potential parallelism in the problem space, but is perhaps not the most efficient.
Especially since most of the action happens on the fringes of the search tree, most goroutines will be stuck waiting on a reply.
Relying on goroutines also makes translating this approach to languages that do not offer a similar construct much harder.
Using a bounded pool of threads could be considered instead.
We return to this idea in Section \ref{workers}.

\subsection{Channels} \label{channels}

Channels are a primitive in Go that allows for many-to-many communication between different routines.
Channels can be closed, which indicates no more messages will be published on it.
Checking whether a channel is closed can only be done by attempting a read.
Writing to a closed channel is a program-halting error in Go.
Channels can be buffered or unbuffered.
Reading from and writing to a channel is a blocking operation if the buffer is full.

Streams are bidirectional and represented by two channels: one for requests and one for responses (the \emph{inbox}).
The actual output stream is the inbox of the parent that the process managing the stream publishes to, which can change during evaluation.
This pattern was chosen to mimic the actor model, in which each process only has an inbox and communicates by sending messages to other processes.
Our implementation relies on parents closing their children's request channels, signalling that no more requests will come and work can halt.
Processes managing streams are only responsible for the lifecycle of their inbox.
Messages clear up this interface and abstract over the Go internals so this approach can be embedded in different languages.
Closing a channel is replaced by sending a \emph{done}-message instead, with the same effect.
A full implementation of \texttt{mplus} in this style is included in the Appendix.

\subsection{Context cancellation}

The last Go primitive for concurrency is the \texttt{select} statement, which provides a safe way to listen to results from multiple channels.
A typical usage is for processes to pass down a context object to subprocesses, which contains a dedicated channel for signals to stop execution.
Whenever a process blocks on read, it uses \texttt{select} to also listen to the context signalling execution should be halted.
This allows processes to be cleaned up even if they were otherwise performing a blocking read.
Since we already have two-way communication between processes over streams, we have opted to use an explicit done-message instead.
With thousands of goroutines, \texttt{select} incurs considerable overhead in the scheduler, which we can now avoid.

The reason we mention it still is that context cancellation patterns can provide inspiration for improving our own closing mechanism.
Especially when working with infinite streams, defining exactly when a subprocesses should give up on evaluation can be very useful.
Typical examples include deadlines and hierarchical contexts.
Before deciding to not use context cancellation in our implementation, our \texttt{conj\_sce} for example used context deadlines to specify the amount of time we are willing to wait on an infinite stream to be productive before giving up and returning an empty result.

\subsection{Passing substitutions}

In concurrentKanren, multiple processes are constantly sharing state by passing messages.
The main component of state is a substitution, which is a complex data structure that can be expanded upon by child processes.
Doing so concurrently is not safe by default: updating the same variable might lead to a clash.
We could copy the entire substitution whenever we pass it such that the copy can safely be modified, but doing so would be very slow.
Our solution is to use immutable AVL trees \cite{avl1963} with structure sharing.
This is different from representations discussed in \cite{Bender2009}, however concurrency is a new consideration.
Nodes in the AVL tree represent key-value pairs in the substitution, associating a variable with an expression.
An entire tree can be shared as a single pointer, since we will never make an update to it after creating it.
Updates instead copy and modify only a few nodes of the original tree, with the new nodes referring to most of the old tree's structure.
Insertions create a minimal amount of new nodes, starting where the update fits in the new tree and moving up to the root.
Compared to copying the entire tree on each update, this reduces the copying overhead from O($n$) to O(log $n$).

\subsection{Worker pool} \label{workers}

So far we have focused on an implementation that relies on Go primitives.
Each stream has a dedicated goroutine managing it.
As we will see in Section \ref{results}, this is not sustainable: spawning many goroutines gets very inefficient.
We can limit the number of goroutines, in the process decoupling our implementation to the specifics of Go, by using a thread pool pattern.
In Go this pattern is usually referred to as the \emph{worker pool} pattern.
It requires us to define query evaluation in miniKanren not as actors maintaining streams, but as pieces of work that suspend and resume.
We define goals as structs, encapsulating the same values we previously captured in a function closure.
When a goal is applied to a state, it now spawns a piece of work that is enqueued in a global queue.
A limited number of worker routines continuously pick up work from the queue and put new work on the queue.

\begin{verbatim}
    func mplus(str, str1, str2 stream) {
        suspendOnRequest(str,  func(sender stream, done bool) {		
            str.request(str1)						
            suspendOnResult(str, func(st state, ok bool) {						
                if !ok {
                    str.forward(str2)						// forward results
                    return
                }
                str.publish(res)											// publish new result
                mplus(str, str2, str1)					// continue with interleaving
            })
        })
    }
\end{verbatim}

Compare the earlier approximate \texttt{mplus} implementation of Section \ref{searchnodes} with the version above.
Work is defined as a suspension point and a continuation.
Suspension either happens waiting on a request or a result from another stream.
While previously we relied on the Go scheduler to handle this for us using channels, we now have to rely on our own task queue implementation to suspend and resume work as messages become available.
This approach can be replicated using any programming language that supports threading.
With a bit more extra work, our approach is now fully language-agnostic.
Standard context cancellation patterns are also viable again, since the number of goroutines is bounded.

\section{Results} \label{results} 

\begin{figure}
\centering
\begin{subfigure}{.5\textwidth}
  \centering
\resizebox{\linewidth}{!}{%
\begin{tikzpicture}
\begin{axis}[
    xlabel={num},
    ylabel={time (s)},
    xmin=0, xmax=600000,
    ymin=0, ymax=180,
    xtick={0,100000, 200000, 300000, 400000, 500000, 600000},
    ytick={0,60,120,180,240,300,360,420,480},
    legend pos=north west,
    ymajorgrids=true,
    grid style=dashed,
]

\addplot
    coordinates {
     (10000,2.1)
    (100000,16)
    (200000,36)
    (300000,77)
    (400000,154)
    (500000,85)
    (600000,361)
    };

\addplot
    coordinates {
     (10000,1.1)
    (100000,9.8)
    (200000,19.6)
    (300000,40.3)
    (400000,56.5)
    (500000,45.8)
    (600000,262)
    };

\addplot
    coordinates {
     (10000,1.3)
    (100000,10.2)
    (200000,21)
    (300000,42.9)
    (400000,56.8)
    (500000,48)
    (600000,175)
    };

\addplot
    coordinates {
     (10000,1.1)
    (100000,8.9)
    (200000,17.5)
    (300000,36)
    (400000,50.4)
    (500000,43.8)
    (600000,165)
    };

\addplot
    coordinates {
     (10000,1)
    (100000,7.9)
    (200000,16.2)
    (300000,32.2)
    (400000,40.5)
    (500000,41.4)
    (600000,132)
    };

\legend{GOMAXPROCS=2, GOMAXPROCS=4, GOMAXPROCS=6, GOMAXPROCS=8, GOMAXPROCS=10}
    
\end{axis}
\end{tikzpicture}
}
  \caption{Message-based benchmark}
  \Description{A line plot of the message-based data found in table \ref{tab:res}}
\end{subfigure}%
%
\begin{subfigure}{.5\textwidth}
  \centering
\resizebox{\linewidth}{!}{%
\begin{tikzpicture}
\begin{axis}[
    xlabel={num},
    ylabel={time (s)},
    xmin=0, xmax=600000,
    ymin=0, ymax=180,
    xtick={0,100000, 200000, 300000, 400000, 500000, 600000},
    ytick={0,60,120,180,240,300,360,420,480},
    legend pos=north west,
    ymajorgrids=true,
    grid style=dashed,
]

\addplot
    coordinates {
    (10000,2.174)
    (100000,23.288)
    (200000,52.199)
    (300000,110.620)
    (400000,125.872)
    (500000,98.662)
    (600000,674.943)
    };

\addplot
    coordinates {
     (10000,2.121)
    (100000,20.079)
    (200000,46.654)
    (300000,98.139)
    (400000,108.632)
    (500000,87.249)
    (600000,479.773)
    };

\addplot
    coordinates {
     (10000,2.3)
    (100000,23)
    (200000,50)
    (300000,104)
    (400000,113)
    (500000,92)
    (600000,234)
    };

\addplot
    coordinates {
     (10000,2.3)
    (100000,23)
    (200000,51)
    (300000,105)
    (400000,114)
    (500000,93)
    (600000,143)
    };

\addplot
    coordinates {
     (10000,2.331)
    (100000,23)
    (200000,50)
    (300000,105)
    (400000,113)
    (500000,93)
    (600000,140)
    };

\legend{GOMAXPROCS=2, GOMAXPROCS=4, GOMAXPROCS=6, GOMAXPROCS=8, GOMAXPROCS=10}
    
\end{axis}
\end{tikzpicture}
}
  \caption{Worker pool benchmark}
  \Description{A line plot of the worker pool-based data found in table \ref{tab:res}}
\end{subfigure}%
\begin{subfigure}{.5\textwidth}
  \centering
\resizebox{\linewidth}{!}{%
\begin{tikzpicture}
\begin{axis}[
    xlabel={num},
    ylabel={time (s)},
    xmin=0, xmax=1000000,
    ymin=0, ymax=500,
    xtick={0,100000, 200000, 300000, 400000, 500000, 600000, 700000, 800000, 900000, 1000000},
    ytick={0,60,120,180,240,300,360,420,480},
    legend pos=north west,
    ymajorgrids=true,
    grid style=dashed,
]

\addplot
    coordinates {
     (10000,1)
    (100000,8)
    (200000,17)
    (300000,35)
    (400000,45)
    (500000,43)
    (600000,136)
    (700000,212)
    (800000,475)
    (900000,420)
    (1000000,451)
    };

\addplot
    coordinates {
     (10000,2)
    (100000,14)
    (200000,31)
    (300000,64)
    (400000,67)
    (500000,56)
    (600000,140)
    (700000,132)
    (800000,147)
    (900000,126)
    (1000000,124)
    };

\addplot
    coordinates {
     (10000, 1)
    (100000, 9)
    (200000, 22)
    (300000, 39)
    (400000, 62)
    (500000, 47)
    (600000, 326)
    (700000, 142)
    (800000, 529)
    (900000, 215)
    (1000000, 320)
    };

\addplot
    coordinates {
     (10000, 1)
    (100000, 9)
    (200000, 21)
    (300000, 48)
    (400000, 48)
    (500000, 37)
    (600000, 122)
    (700000, 113)
    (800000, 112)
    (900000, 101)
    (1000000, 86)
    };

\legend{message-based, worker pool, single-threaded, faster-minikanren}
    
\end{axis}
\end{tikzpicture}
}
  \caption{Comparison}
  \Description{A line plot of the message-based data found in table \ref{tab:res}}
\end{subfigure}%
\caption{Benchmark plots of message-based and worker pool implementations}
\label{fig:test}
\end{figure}
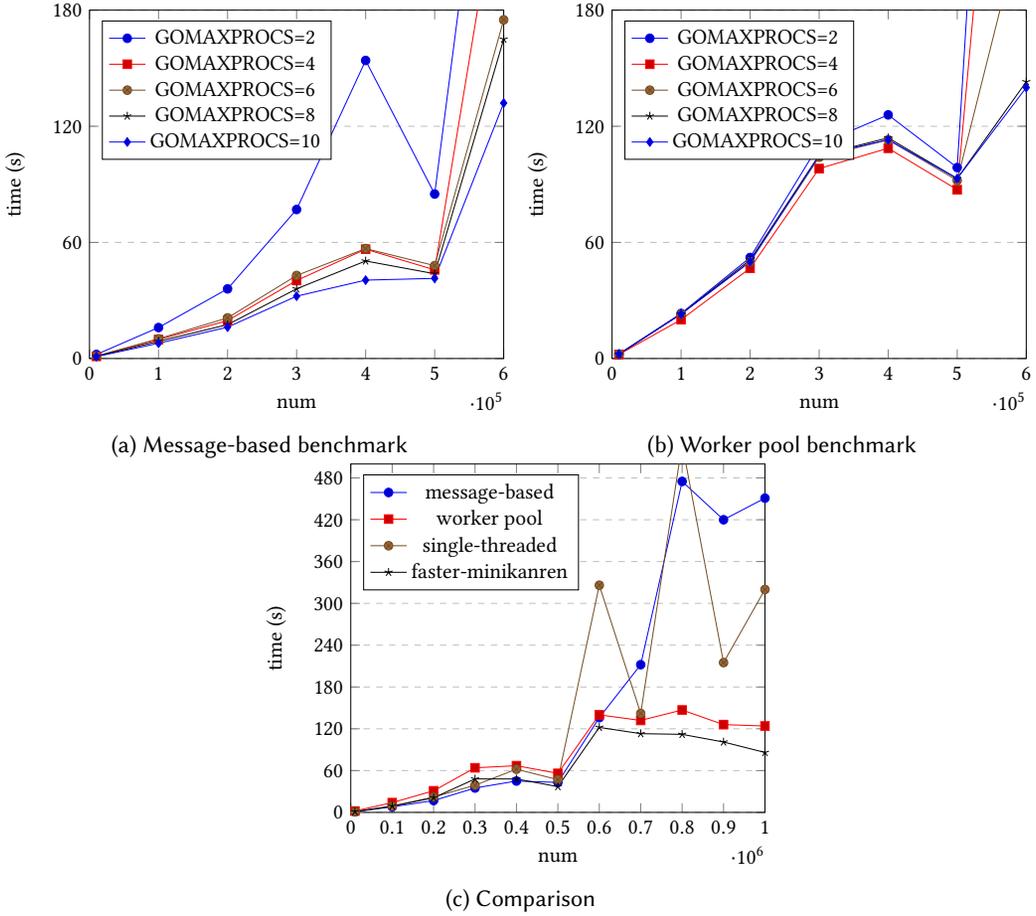

We investigate the execution times of the message-based implementation and the worker pool version.
To do so we use a benchmark query that, while somewhat contrived, has a single parameter that scales up the complexity of the associated search tree.
Consider the following query:

\begin{verbatim}
> (define (sums-to-n q n) (conj (equalo q `(,x ,y)) (plusO x y n)))
> (run* (q) (sums-to-n q num)))
\end{verbatim}

where \texttt{num} is an Oleg-numeral \cite{Kiselyov2008}.
This query returns all the ways that two natural numbers can sum up to \texttt{num}, which is \texttt{num} $+ 1$.
The benchmark was run on a MacBook Pro with 16GB memory and an Apple M4.
Times are measured using the builtin \texttt{time} command.
Queries are repeated five times and results averaged.
Figure \ref{fig:test} shows the results: \ref{fig:test}a and \ref{fig:test}b show the results for message-based and worker pool implementations respectively, ran using a varying amount of cores available.
In Go this can be achieved by setting the \texttt{GOMAXPROCS} system variable.
Figure \ref{fig:test}c compares the two implementations, both utilising the maximum number of cores available and using \texttt{disj\_conc} for disjunctions over more than two goals.
We include \texttt{faster-minikanren} as a reference, and a simple single-threaded Go implementation as a baseline.\\

We can see that both implementations benefit from more cores, especially when the problem size is large.
The comparison shows that at some point the message-based approach incurs a heavy penalty from overhead, while the worker pool implementation manages to avoid it.
Its initial overhead cost is a bit higher but scales a lot better.
Our worker pool implementation follows the reference implementation quite closely, slightly faster on smaller benchmarks but a little slower for non-trivial cases.
The single-threaded baseline shows heavy variance, even beating \texttt{faster-minikanren} on certain larger cases.
We cannot fully explain why, but it is clear that generally it performs much worse than the rest.
A similar trend but smaller is visible in all graphs: some larger cases take less time than smaller ones.
We believe this is due to the structure of the search tree associated with our benchmark, but have not investigated in depth.
Notice for example how the reference implementation gets progressively faster, which is not what we would expect.

\section{Conclusions and Future Work}

We have introduced \emph{concurrentKanren} and demonstrated two approaches to implementation.
Our approaches take inspiration from the actor model to focus on message passing between processes.
For both disjunction and conjunction, we have shown goals that make use of concurrency to improve the miniKanren language.
The result is shown to be an improvement over the naive baseline, and competitive with existing standards.\\

We encourage exploration of concurrent miniKanren implementations in other languages.
As mentioned in Section \ref{workers}, we recommend using a reusable worker pool architecture available to any language with threading support.
Finally, our current work is limited to a single-machine setup.
Our focus on passing messages makes the approach well-suited for distributed computing.
Extending it to a distributed environment presents a promising direction for future research.

\begin{acks}
We would like to thank Jason Hemann and Rebecca Swords for their help and feedback, Ron Evans for the motivation, and Walter Schulze for the introduction to miniKanren all those years ago.
\end{acks}

\bibliographystyle{ACM-Reference-Format}
\bibliography{biblio}

\appendix

\section{Full mplus}
\begin{verbatim}
     func mplus(str, str1, str2 stream) {
          parent, done := str.getRequest()
          if done {
                str1.sendDone()
                str2.sendDone()
                str.close()
               return
          }
          mplus_(parent, str, str1, str2)
     }

     func mplus_(parent, str, str1, str2 stream) {
          str.request(str1)
          rec, ok := str.receive()
          if !ok {
               panic("mplus tried to read from closed channel")
          }
          switch {
          case rec.isState():
               str.sendState(parent, rec.st)
               mplus(str, str2, str1)
          case rec.isStateAndClose():
               str.sendForwardWithState(parent, str2, rec.st)
          case rec.isClose():
               str.sendForward(parent, str2)
          case rec.isForward():
               mplus_(parent, str, rec.fwd, str2)
          case rec.isForwardWithState():
               str.sendState(parent, rec.st)
               mplus(str, str2, rec.fwd)
          case rec.isDelay():
               mplus_(parent, str, str2, str1)
          }
     }
\end{verbatim}

\section{Generalized concurrent disjunction}
\begin{verbatim}
    func disj_conc(goals ...goal) goal {
        return func(st state) stream {
            str := newStream()
            streams := applyAll(goals, st)
            go mplusplus(str, nil, streams)
            return str
        }
    }
    func mplusplus(str stream, buffer []state, active []stream) {
        if len(buffer) == 0 {
            buffer, active = refillBuffer(str, active)
        }
        str.getRequest()
        if len(buffer) == 0 {
            str.close()
            return
        }
        str.publish(buffer[0])
        mplusplus(str, buffer[1:], active)
    }
\end{verbatim}

\section{Conjunction with short-circuit evaluation}
\begin{verbatim}
    func conj_sce(g1, g2 goal) goal {
        return func(st state) stream {
            str := newStream()
            str1 := conj(g1, g2)(st)
            str2 := g2(st)
            go short_circuit(str, str1, str2)
            return str
        }
    }
    func short_circuit(str, str1, str2 stream) {
        str1.requestResult()
        str2.requestResult()
        select {
        case msg <- str1:
            str.publish(msg)
            str.forward(str1)
            return
        case msg, ok <- str2:
            if !ok {
                str.close()
                return
            }
            str.waitOnStrAndForward(str1)
        }
    }
\end{verbatim}

\end{document}